\documentclass[]{aa}
\usepackage{amsmath}
\usepackage{latexsym}
\usepackage{longtable}
\usepackage{natbib}
\usepackage[latin1]{inputenc}
\usepackage{graphics}
\usepackage{epsfig}
\usepackage{graphicx}

\begin{document}

\title{To the center of cold spot with Planck}
\author{V.G. Gurzadyan\inst{1,2}, A.L. Kashin,\inst{2}, H. Khachatryan\inst{2}, E. Poghosian\inst{2},
S. Sargsyan\inst{2}, G. Yegorian\inst{2}}

\institute
{\inst{1} SIA, Sapienza University of Rome, Rome, Italy\\
\inst{2} Center for Cosmology and Astrophysics, Alikhanian National Laboratory and Yerevan State University, Yerevan, Armenia\\
}

\date{Received (\today)}

\titlerunning{CMB}

\authorrunning{V.G.Gurzadyan et al}

\abstract{
The structure of the cold spot, of a non-Gaussian anomaly in the cosmic microwave background (CMB) sky first detected by Vielva et al. is studied using the data by Planck satellite. The obtained map of the degree of stochasticity (K-map) of CMB for the cold spot, reveals, most clearly in 100 GHz band, a shell-type structure with a center coinciding with the minima of the temperature distribution. The shell structure is non-Gaussian at a $4\sigma$ confidence level. Such behavior of the K-map supports the void nature of the cold spot.  The applied method can be used for tracing voids that have no signatures in redshift surveys.}


\maketitle

\section{Introduction}

The cold spot is a non-Gaussianity in the Cosmic Microwave Background (CMB) sky, which since its first detection \cite{V} in Wilkinson Microwave Anisotropy Probe (WMAP) data has attracted much attention in the studies with various statistical descriptors of non-Gaussian features in general, as well as in its interpretation \cite{Cruz,Piet,GS,DS,V10,Val,R,Fern,Co,Ko,Zh,Planck}; and references therein).
  
A void (i.e., an underdense region in the matter distribution) has been discussed as one of the possible interpretations of the properties of the cold spot \cite{In,MN,Ino,Ko}. The results represented below offer additional support to the void interpretation of the cold spot and are based on the study of the stochasticity (randomness) properties in the temperature data of Planck \cite{Planck}. Owing to the hyperbolicity of null geodesics the degree of randomness towards the walls of a void has to be higher than in its inner region \cite{GK_void1,GK_void2}. This property of the cold spot has already been noticed in the WMAP data \cite{G2009}, and as several other sky regions with similar properties have been identified. The randomness of the CMB signal has been studied \cite{GK_KSP} using the Kolmogorov stochasticity parameter  \cite{K,UMN,MMS,FA}, which enables determining the degree of randomness in the given sequence of real numbers. That method enabled us to reveal X-ray galaxy clusters in XMM-Newton data \cite{Xray} and to detect the thermal trust Yarkovsky-Rubincam effect for the satellites measuring the Lense-Thirring effect predicted by general relativity \cite{lageos}. 

Here we show that the Kolmogorov map (K-map) of Planck's 100 GHz data reveal a shell-like structure in the cental region of the cold spot, and this supports its void nature.   

\section{Hyperbolicity}

We consider the behavior of null geodesics in underdense regions (cf. \cite{Z}). The equation of geodesic deviation for perturbed Friedmann-Robertson-Walker metric can be reduced to \cite{GK_void1} 
\begin{equation}
\frac{d^2\ell}{d\lambda^2}+ r\ \ell = 0\ ,
\end{equation}
where $l$ is the deviation scalar,
\begin{equation}
\lambda(z,\Omega_\Lambda,\Omega_m)
=\int_0^z\frac{d\xi}{\sqrt{\Omega_\Lambda +[1- \Omega_\Lambda + \Omega_m \xi]\ (1+\xi)^2}},\
\end{equation}
and $\Omega_\Lambda$ and $\Omega_m$ are the density parameters for the dark energy and matter, respectively.  
For flat FRW metric the scalar $r$ has the form 
\begin{equation}
r = 2\Omega_m\delta_0\ ,
\end{equation}
where 
\begin{equation}
\delta_0\equiv \frac{\delta\rho_0}{\rho_0}\ ,
\end{equation}
and $\delta\rho_0$ is the local density contrast with respect to the mean matter density $\rho_0$.

For voids,
$$\delta_{void}=\frac{\rho_{void}-\rho_0}{\rho_0}$$
is negative, thus ensuring the hyperbolicity of null geodesics; i.e., the low density contrast locally acts as a negative curvature, even in  a flat universe.

\section{K-maps}

The randomness induced by the hyperbolicity of null geodesics can be studied using CMB data and the Kolmogorov stochasticity parameter \cite{K,UMN,MMS,FA}. The latter is defined for $\{X_1,X_2,\dots,X_n\}$ real-valued variables $X$ represented as $X_1\le X_2\le\dots\le X_n$. 
For such a sequence the empirical distribution function $F_n(x)$ is given as 
\begin{eqnarray*}
F_n(x)=
\begin{cases}
0\ , & x<X_1\ ;\\
k/n\ , & X_k\le x<X_{k+1},\ \ k=1,2,\dots,n-1\ ;\\
1\ , & X_n\le x\ ,
\end{cases}
\end{eqnarray*}
and the cumulative distribution function (CDF) is  
$
F(x) = P\{X\le x\}. 
$
The stochasticity parameter $\lambda_n$ is estimated as   
$
\lambda_n=\sqrt{n}\ \sup_x|F_n(x)-F(x)|.
$

According to Kolmogorov's theorem \cite{K},  one has the limit
$
\lim_{n\to\infty}P\{\lambda_n\le\lambda\}=\Phi(\lambda)\ ,
$
for continuous CDF, so that $\Phi(0)=0$, and 
\begin{equation}
\Phi(\lambda)=\sum_{k=-\infty}^{+\infty}\ (-1)^k\ e^{-2k^2\lambda^2}\ ,\ \  \lambda>0\ ,\label{Phi}
\end{equation}
and $\Phi$ is independent on $F$. The interval $0.3\le\lambda_n\le 2.4$ corresponds to the variation scale of $\Phi$ as of
objective measure of degree of randomness.

This approach has been applied for several sequences of the theory of dynamical systems and number theory  \cite{UMN,MMS,FA,Entr} and for extensive modeling of generated systems (see \cite{mod} and references therein). In the applications to CMB in view of its Gaussian feature, the Gaussian CDF was used for analyzing of the temperature sequences.  The obtained Kolmogorov CMB maps (K-maps) for WMAP data readily reveal, for example, the Galactic disk and point sources (galaxies, quasars) that have different randomness (correlation) properties than the cosmological signal \cite{G2009}, thus justifying the efficiency and predictive power of the approach.

The randomness of the cold spot has also been studied for WMAP data using this technique, and the results correspond to features expected for voids, i.e. underdense regions, with the increase in randomness toward their walls \cite{G2009}. Also, a sample of regions has been revealed in the CMB sky, with similar behavior, although less outlined, of the function $\Phi$ vs the radius.

We now analyze the cold spot region using Planck's temperature data (nominal) in HEALPix format \cite{HP}. The computations were performed in the following manner (for details see \cite{G2009}. The temperature data sequences in ball regions of radius 0.5 degree located at step 0.05 degree have been chosen, containing about 950 pixels each for 100 GHz Nside=2048 maps, and for them the function $\Phi$ has been obtained. The stability of the results was tested by varying the input parameters, most importantly the step (varying it e.g. within 0.01 - 0.05 degree) and the radius (see also \cite{G2009}), thus ensuring the robustness of the results.

First, using the temperature cuts for the cold spot region, we determined the minima of the 100 GHz temperature plots, possessing the coordinates (210.04$^\circ$, -56.26$^\circ$), as shown in Fig.1.  Then, in Fig. 2 we represent the 100 GHz temperature Nside=2048 map (in $\mu K$ scale, upper plot) and the corresponding K-map for the cold spot region, with denoted temperature minima location. A shell-like structure with a 0.6 degree  radius of external boundary is visible in the K-map, which is absent in the temperature map and clearly surrounds the `center' defined by temperature minima. The same shell structure is also available in 143 GHz K-map (Fig.3), again without a trace in its temperature map, but has no such outlined signature in Nside=2048 70 GHz K-map (Fig.4). To quantify the confidence level of the shell, we used the same algorithm to estimate the behavior of $\Phi$ vs the radius for 200 regions of a Gaussian map with $T_{mean}=27.47 \mu K$ and $\sigma=119.99\, \mu K$ of the Planck's 100 GHz map of the northern sky without the Galactic disk region ($|b| < 20$ degree). As seen from Fig. 5, the jump of $\Phi$ of the cold spot yields over 4$\sigma$ [(0.95-0.14)/0.2].

\begin{figure}[ht]
\centerline{\epsfig{file=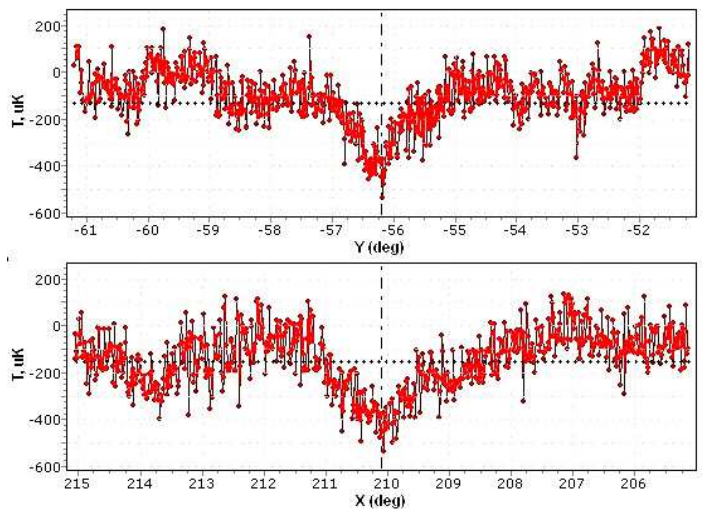,width=0.5\textwidth}} \vspace*{8pt}
\caption{Temperature run vs latitude and longitude for the cold spot region for Planck's 100 GHz map.}
\end{figure}

\begin{figure}[ht]
\centerline{\epsfig{file=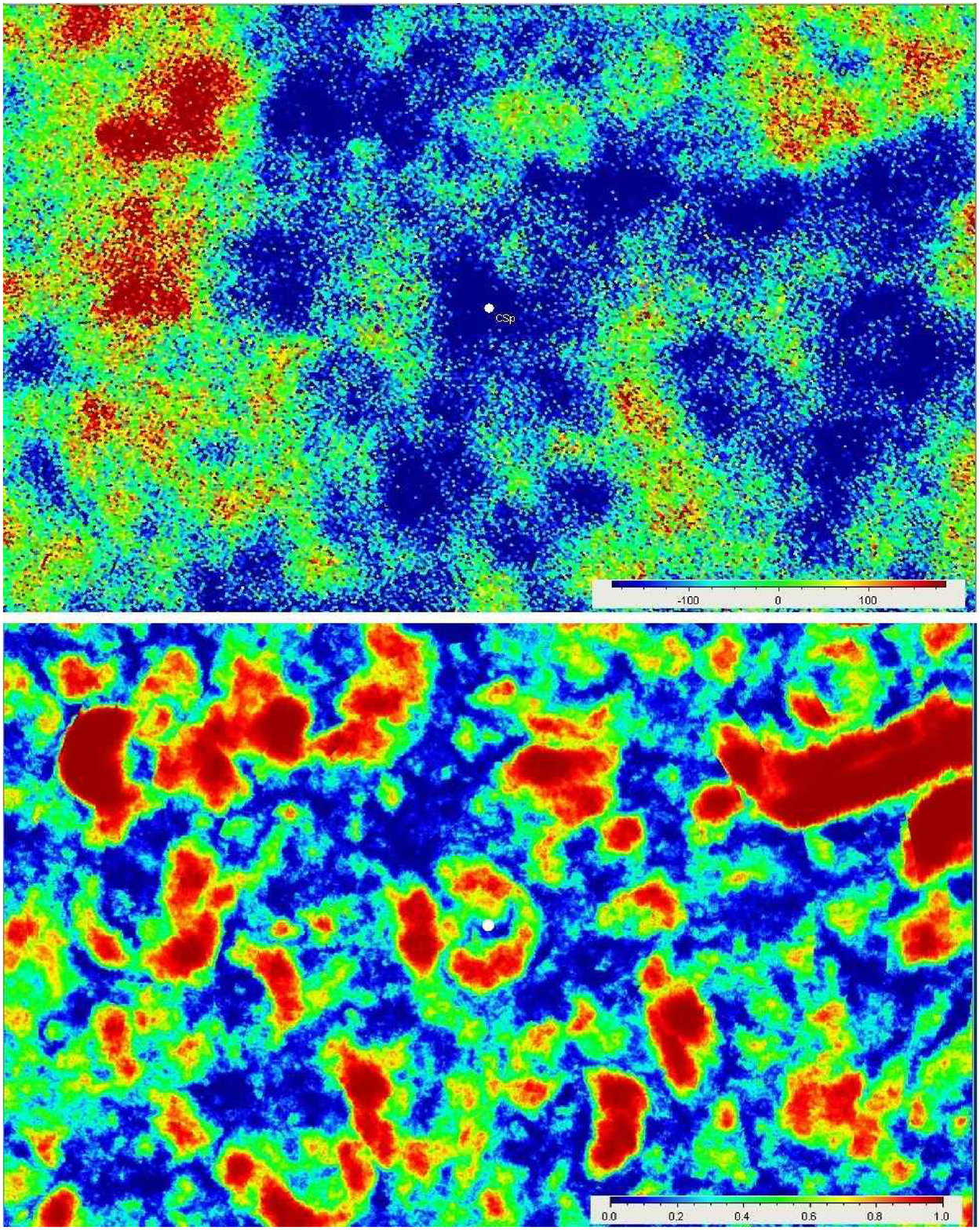,width=0.5\textwidth}} \vspace*{8pt}
\caption{100 GHz temperature map (above) and its K-map. The denoted point of coordinates (210.04$^\circ$, -56.26$^\circ$) corresponds to the minima of the temperature run in Fig.1.}
\end{figure}

\begin{figure}[ht]
\centerline{\epsfig{file=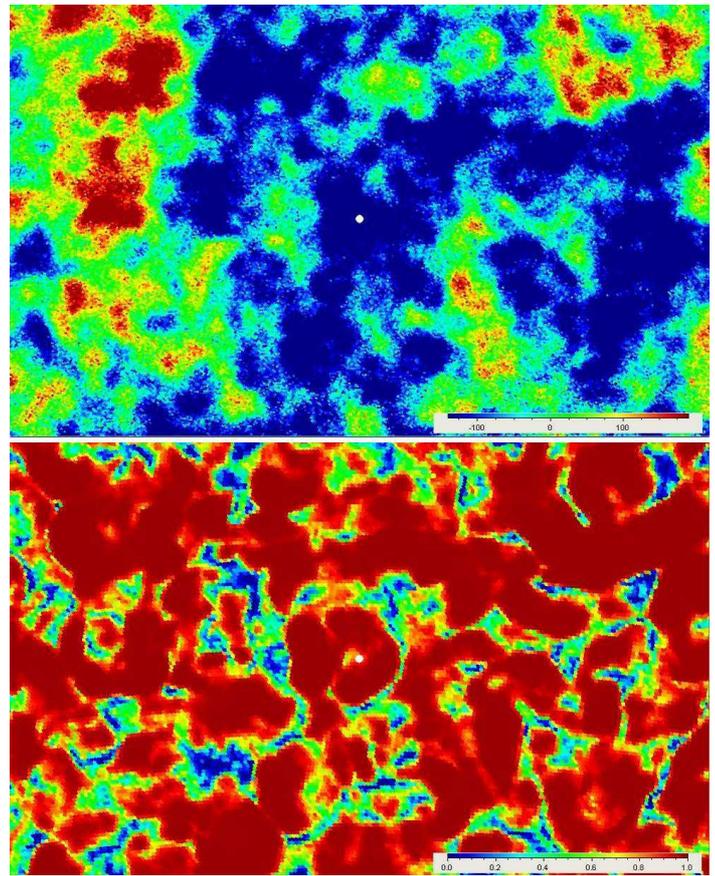,width=0.5\textwidth}} \vspace*{8pt}
\caption{Same as in Fig.2 but for 143 GHz.}
\end{figure}

\begin{figure}[ht]
\centerline{\epsfig{file=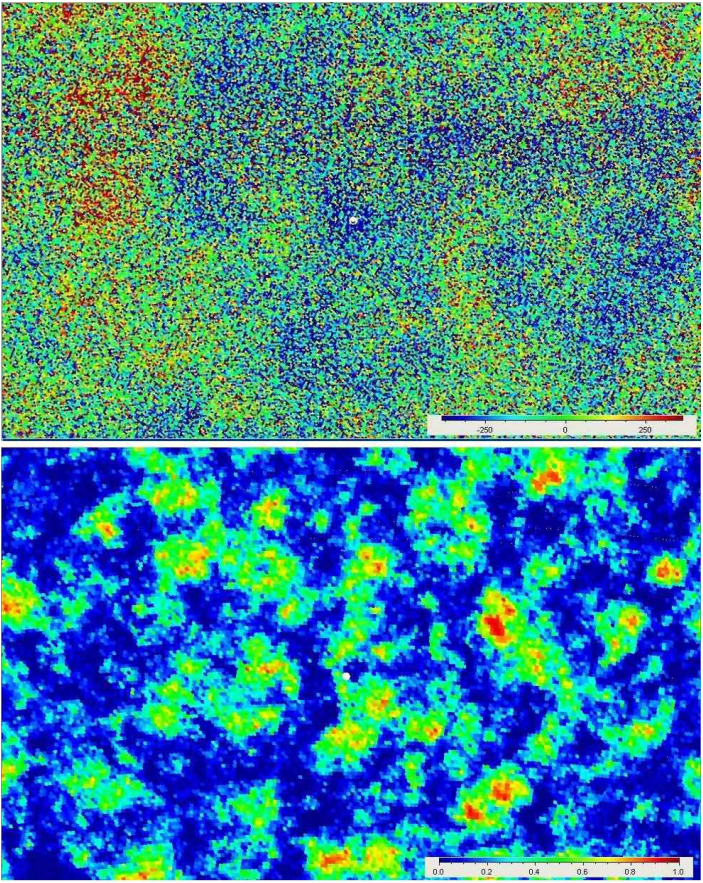,width=0.5\textwidth}} \vspace*{8pt}
\caption{Same as in Fig.2 but for 70 GHz.}
\end{figure}

\begin{figure}[ht]
\centerline{\epsfig{file=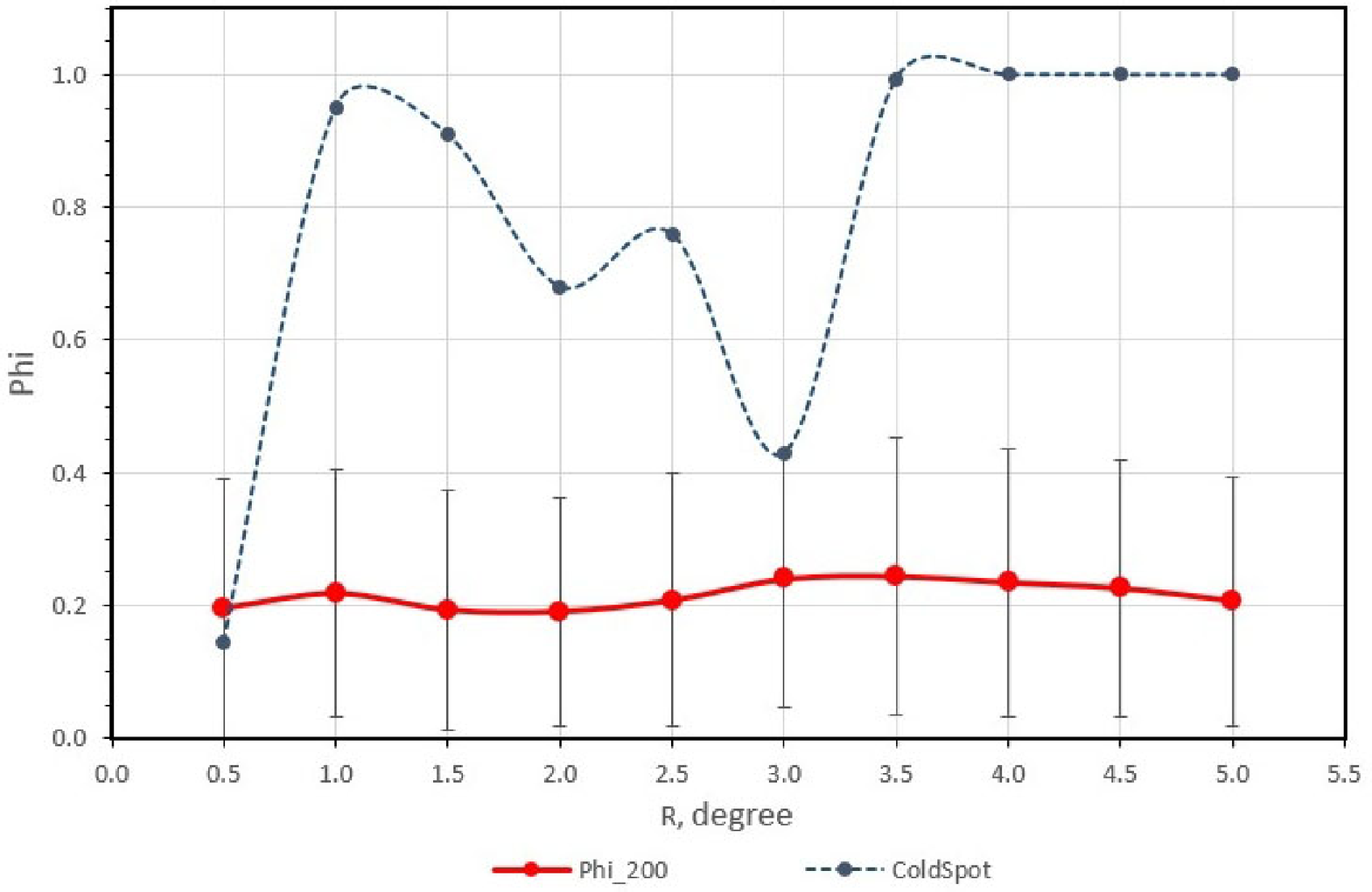,width=0.5\textwidth}} \vspace*{8pt}
\caption{Dependence of $\Phi$ vs the radius for the cold spot for 100 GHz data and for 200 regions of simulated Gaussian map of the parameters of the 100 GHz map (more in the text).}
\end{figure}

\section{Conclusions}

We have studied the structure of the cold spot using the Planck CMB temperature maps and obtained the corresponding K-maps, i.e., the distribution of the degree of randomness in that region. The quality of Planck data enabled us to reveal in the 100 GHz K-map a shell-like structure, the center of which does coincide with the position of the temperature minima. Such a shell structure is also clear in 143 GHz K-map but has no such visible trace in the 70 GHz K-map (Nside=2048). 

The shell-type distribution of the degree of randomness is expected at the hyperbolicity of null geodesics induced by voids. The obtained structure in the K-map does therefore support the void nature of the cold spot. It is worth mentioning that the universality of the applied method enables a void to be traced independently of whether its walls are outlined well by luminous matter or mostly contain dark matter,  hence have no clear correlations with available galaxy redshift surveys \cite{Br,Gr}. The void of the cold spot, therefore, may need more refined alternative means for study. The importance of the presence of voids with determined parameters is obvious for understanding the formation of large scale structure formation in the universe.        

\section{Acknowledgments}

We acknowledge the useful comments by the referee and the use of Planck data (http://pla.esac.esa.int/pla/aio/planckProducts).

\end{document}